\documentstyle[12pt]{article}
\def\be{\begin{equation}}
\def\ee{\end{equation}}
\def\bea{\begin{eqnarray}}
\def\eea{\end{eqnarray}}
\def\l{\label}
\def\c{\cite}

\begin{document}
\begin{titlepage}

\flushright{To Appear: {\em Phys. Lett.} {\bf B}}

\vspace{2cm}

\begin{center}
\Large
{\bf Wavefunctions for  highly anisotropic homogeneous cosmologies}

\vspace{2cm}

\normalsize

\large{James E. Lidsey$^{1}$}

\normalsize
\vspace{1cm}

{\em Astronomy Unit, School of Mathematical Sciences, \\
Queen Mary \& Westfield, Mile End Road, LONDON E1 4NS, UK}

\vspace{3.5cm}

\end{center}

\baselineskip=24pt
\begin{abstract}
\noindent

The canonical quantization of  homogeneous cosmologies is considered
in the high anisotropic limit.
Exact wavefunctions are found in this limit when the momentum
constraints are reduced at the classical level. Lorentzian
solutions that represent
tunnelling from classically forbidden regimes are identified.
Solutions to the modified Wheeler-DeWitt equation are also
found for the vacuum Bianchi IX
model when a quantum reduction of the momentum constraints is considered.

\end{abstract}

\vspace{1cm}

\vspace{1cm}

\small{$^1$Electronic address: jel@maths.qmw.ac.uk}

\normalsize

\end{titlepage}

\setcounter{equation}{0}


Quantum cosmology applies the concepts of quantum mechanics to the
Universe as a whole. The majority of studies in this field  invoke the
Dirac quantization  procedure \c{dirac}.
In this approach the wavefunction of the Universe
is annihilated by the operator versions of the Hamiltonian and momentum
constraints \c{wheeler}. However, it is not known how to solve
these constraints  in full generality. Indeed, the configuration
space of the wavefunction is infinite-dimensional.
In practice, therefore, one invokes the
`minisuperspace' approximation and applies the procedure only to
those cosmologies that represent homogeneous solutions to Einstein's field
equations \c{wheeler,misner}.
These are the Bianchi models and the Kantowski-Sachs Universe.
Hence, the problem is reduced to a quantum mechanical system with a finite
number of degrees of freedom.

It is not clear whether
the results derived from  this minisuperspace quantization
represent a valid approximation to the  full theory of quantum gravity
\c{KR1989,GHM1991}. However, the main justification for this
approach is that it  provides a solvable framework in which
the problems associated with  quantum cosmology may be addressed.
For example, there are difficulties in extracting physical predictions
from the wavefunction and there are also problems with defining
a Hilbert space of states. (For a review see, e.g., Ref \c{H1990}).

One may adopt one of two approaches
when performing a canonical quantization of homogeneous  Universes.
The simplest method is
to reduce the momentum constraints at the classical level before quantization
and most investigations  to date have adopted this view. The
Wheeler-DeWitt equation represents the quantum
Hamiltonian  constraint and
governs the evolution of the wavefunction on minisuperspace \c{wheeler}.
However, a more rigorous and consistent  method  is to reduce the
momentum constraints at the quantum level.  In this way
all constraints are treated in a similar fashion.
These two
approaches are not necessarily equivalent and can lead to different results
\c{AS1991,HW1995}.

In either case, however, very few exact wavefunctions have been found to date.
The purpose of this letter is to derive new families  of  solutions
to the quantum  constraints of the Bianchi class A models. We consider
wavefunctions on a region of minisuperspace corresponding
 to the high anisotropic limit of these models.

We first investigate the
approach whereby the momentum constraints are reduced prior to quantization.
We show that the resulting Wheeler-DeWitt equation may be written in the form
of the unit-mass Klein-Gordon equation. We then solve  the
Wheeler-DeWitt equation for the vacuum Bianchi IX model when
a full quantum reduction of the momentum constraints is performed.

The homogeneous Bianchi space-times have
a topology $R\times G_3$, where $G_3$ is a
three-dimensional Lie group of isometries
transitive on space-like three-dimensional
orbits \c{RS1975}. The world-interval of the space-time
is
\be
\l{interval}
ds^2 =-dt^2 +h_{ab} \omega^a \omega^b , \qquad a,b=1,2,3,
\ee
where the metric $h_{ab}$ on the surfaces of homogeneity is a function of $t$
alone and
$\omega^a$ are one-forms. The isometry of the
three-surface is determined  by the structure constants
${C^a}_{bc}$ of the Lie algebra of $G_3$. These may be decomposed as
${C^a}_{bc} = m^{ad} \epsilon_{dbc} +{\delta^a}_{\left[ b \right.}
a_{\left. c \right] }$, where $a_c \equiv {C^a}_{ac}$ and $m^{ab}$
is symmetric \c{EM1969}.
The Jacobi identity ${C^a}_{b \left[ c \right. } {C^b}_{\left. de \right]
} =0$ is satisfied if and only if  $a_b$ is transverse  to $m^{ab}$, i.e.
$m^{ab} a_b =0$. The Lie algebra belongs to the Bianchi
class A if $a_b=0$ and to the class B if $a_b\ne 0$ \c{EM1969}. It is well
known that the evolution of  class B models cannot
be described in terms of  a standard Hamiltonian treatment.
This is related to the fact that these models cannot admit a
spatially compact topology \cite{AS1991}.
Consequently,
we restrict our attention to the class A. This consists
of Bianchi types I, II, ${\rm VI_0}$, ${\rm VII_0}$, VIII and IX.
The Lie algebra of each  type is uniquely  determined up to
isomorphisms by the rank and signature of $m^{ab}$.

We will assume that the matter source is a single, massless,
minimally coupled scalar field $\phi$.
The classical dynamics of these cosmologies is determined by the
time-space and time-time components of the Einstein field
equations. These may be expressed as the momentum and Hamiltonian
constraints, respectively. The former is given by
\be
\l{classmom}
P_d  \equiv {\pi^a}_c m^{cb} \epsilon_{dab} =0 ,
\ee
where the conjugate momentum variable $\pi^{ab}$
represents the integral of the momentum
density over the spatial hypersurfaces.
The Hamiltonian constraint, on the other hand, takes the form \c{W1983}
\be
\l{classham}
\left( \pi^{ab}\pi_{ab} -\frac{1}{2} \pi^2
\right) +{\cal{V}}^2  \left( m^{ab} m_{ab} -\frac{1}{2} m^2 \right)
+\frac{1}{24} \pi^2_{\phi} =0 ,
\ee
where $\pi_{\phi}$ is the momentum conjugate to
the scalar field, ${\cal{V}}$ represents the volume of space and
indices   are raised and lowered with $h^{ab}$ and
$h_{ab}$, respectively.

At the classical level, the
momentum constraints (\ref{classmom})
imply that $\pi^{ab}$, $m^{ab}$ and $h_{ab}$ may be simultaneously
diagonalized on a given spatial hypersurface of constant $t$. It then
follows directly from the field equations
that these quantities are diagonal on all other  hypersurfaces.
Hence, one may diagonalize the variables at the classical
level without loss of generality and
the momentum constraints therefore become trivial.

\begin{table}
\begin{center}
\begin{tabular}{||c||c|c|c|c|c|c||} \hline \hline

Type & I & II & ${\rm VI}_0$ & ${\rm VII}_0$ & VIII & IX\\
\hline
& & & & & & \\
$m^{ab}$ & $(0,0,0)$ & $ (1,0,0)$ & $(1,-1,0)$ & $(1,1,0)$ & $(1,1,-1)$ &
$(1,1,1)$\\
& & & & & & \\
\hline \hline
\end{tabular}
\end{center}
\vspace{.35in}

\small{Table 1: The diagonal components  of
$m^{ab}$ for each Bianchi type in the class A.}

\end{table}

In this case one may choose a basis where each diagonal component
of $m^{ab}$ is $\pm 1$ or $0$.  The six cases  that constitute the
Bianchi class A are  shown in Table 1.
The three-metric is then written as
\be
\l{3metric}
h_{ab}(t)= e^{2\alpha (t)} \left( e^{2\beta (t)} \right)_{ab}
\ee
where the matrix $\beta_{ab} \equiv {\rm diag} \left[ \beta_+ +
\sqrt{3}\beta_-, \beta_+ -\sqrt{3}\beta_- , -2\beta_+ \right]$
represents the degree of anisotropy in these models.
The parameter $e^{\alpha}$ may be viewed as  an averaged
scale factor of  the Universe.
The Hamiltonian constraint (\ref{classham}) then reduces to
\be
\l{classham1}
-\pi^2_{\alpha} +\pi^2_{\beta_+} + \pi^2_{\beta_-} +\pi^2_{\phi}
+  U =0
\ee
where $\pi_j$ $(j=\alpha, \beta_{\pm} ,\phi )$ are the conjugate momenta,
\bea
\l{superpot}
U=\frac{1}{3} \left[ \left( m_{11}h_{11} \right)^2 +
\left( m_{22}h_{22} \right)^2  + \left( m_{33}h_{33} \right)^2 \right.
\nonumber \\
\left.
-2 m_{11}m_{22}h_{11}h_{22} - 2 m_{11}m_{33} h_{11}h_{33}
-2m_{22}m_{33}h_{22}h_{33} \right]
\eea
represents the `superpotential' and
$m^{ab} =   {\rm diag} \left[ m_{11},m_{22},m_{33} \right]$
 \c{HW1995}. (The
direct dependence of the superpotential  on ${\cal{V}}$ is eliminated
by performing a linear translation on $\alpha$). The superpotential for the
Bianchi type I cosmology vanishes identically and
we shall not consider this model further.

Quantization of these cosmologies follows
 by imposing the algebra $[j, \pi_j ]_- =i$.
The Hamiltonian constraint (\ref{classham1}) is then promoted
to an operator that annihilates the state vector $\tilde{\Psi}$ of the
Universe. The result is the Wheeler-DeWitt equation on minisuperspace:
\be
\l{WDW}
\left[ \frac{\partial^2}{\partial \alpha^2} -\frac{\partial^2}{\partial
\beta_+^2} -\frac{\partial^2}{\partial \beta_-^2} +2p
\frac{\partial}{\partial \alpha} - \frac{\partial^2}{\partial \phi^2}
+U \right] \tilde{\Psi} =0 ,
\ee
where the constant $p$
accounts for ambiguities that arise in the operator ordering
\cite{HH1983}.

It is convenient to rescale the wavefunction by $\tilde{\Psi} = \Psi
e^{iq\phi -p \alpha}$,
where $q$ is an arbitrary,  complex constant.
Eq. (\ref{WDW}) then simplifies to
\be
\l{WDW1}
\left[ \frac{\partial^2}{\partial \alpha^2} -\frac{\partial^2}{\partial
\beta_+^2} -\frac{\partial^2}{\partial \beta_-^2}   +q^2 -p^2
+U \right] \Psi =0 .
\ee
A unified exact solution to this equation can be found
for all Bianchi types if we choose the factor ordering
$p^2 =9+q^2$. It can be verified by direct
substitution that the wavefunction has the form
\be
\l{susywave}
\tilde{\Psi}= e^{(3-p) \alpha +iq\phi} e^{-I} ,
\ee
where
\be
\l{action}
I=\pm \frac{1}{6} m^{ab}h_{ab}
\ee
and summation over indices is implied.
This solution generalizes the exact solution found previously
for the type IX model \c{lidsey}.
In the vacuum case $(q=0)$, the function $I$ is itself a solution
to the Euclidean Hamilton-Jacobi equation.  It
therefore  represents a Euclidean action for these cosmologies.

In general, an oscillating solution to the Wheeler-DeWitt equation
may be interpreted as a classically allowed, Lorentzian geometry
\c{H1984}. A cosmological singularity then arises when the
wavefunction undergoes an infinite number of oscillations.
On the other hand, a classically forbidden
Euclidean geometry corresponds to a non-oscillatory,
 exponential wavefunction. In particular, if the wavefunction
is suitably damped at large three-geometries
and remains regular when the spatial metric degenerates, it may be interpreted
as a quantum wormhole \c{HP1990}.
Solution (\ref{susywave}) is an example of this latter type of wavefunction
and therefore represents an anisotropic
 quantum wormhole.

Further unified solutions to Eq. (\ref{WDW1}) can be found
in the limit where  $ \beta_{\pm}  \gg 1$.
This region of minisuperspace  is interesting because quantum
fluctuations of  the gravitational field in  the early Universe  may have
introduced considerable  anisotropy  into the initial conditions.  However,
observations indicate that the present Universe is highly isotropic.
It is  possible, therefore,  that a better understanding of this
high anisotropic regime  may provide insight into how the
Universe evolved into its current state.

It follows that
 $h_{22}/h_{11} \ll 1$ and $h_{33}/h_{11} \ll 1$ when $\beta_{\pm} \gg 1$.
The first term in Eq. (\ref{superpot})
therefore dominates the superpotential.
Formally, this is equivalent to choosing $m_{11}=1$
and $m_{22}=m_{33}=0$. Hence, this particular limit of the
Bianchi class A may be investigated by considering the type II model.
The Wheeler-DeWitt equation (\ref{WDW1}) simplifies  to
\be
\l{WDW2}
\left[ \frac{\partial^2}{\partial \alpha^2} -\frac{\partial^2}{\partial
\beta_+^2} -\frac{\partial^2}{\partial \beta_-^2} +q^2 -p^2
+ \frac{1}{3} e^{4\alpha +4\beta_+ +4\sqrt{3} \beta_-}  \right] \Psi =0
\ee
and may be solved by defining a new wavefunction
\be
\l{new}
\Phi (\alpha ,\beta_{\pm} ) \equiv \Psi (\alpha ,\beta_{\pm} )
e^{- r( \sqrt{3}\beta_+ - \beta_- )/2 } ,
\ee
where
$r^2\equiv  q^2 -p^2$. Substitution of this ansatz into Eq. (\ref{WDW2})
implies that
\be
\l{eq}
\left[ \frac{\partial^2}{\partial \alpha^2} -\frac{\partial^2}{\partial
\beta_+^2} -\frac{\partial^2}{\partial \beta_-^2} -\sqrt{3}r
\frac{\partial}{\partial \beta_+} +r\frac{\partial}{\partial \beta_-}
+\frac{1}{3} e^{4\alpha +4\beta_+ +4\sqrt{3} \beta_-}  \right] \Phi =0 .
\ee

We now perform  a change of variables to $\{ u,v,w \}$, where
\bea
\l{u}
u\equiv \frac{1}{12} e^{6\alpha +3\beta_+ +3\sqrt{3} \beta_-}
\nonumber \\
v \equiv \frac{1}{12} e^{-2\alpha +\beta_+ +\sqrt{3} \beta_-}
\eea
and $w=w(\alpha ,\beta_{\pm})$
is an arbitrary, twice continuously differentiable
 function of the minisuperspace coordinates.
We then search for wavefunctions  that are independent of this variable.
In this case, Eq. (\ref{eq}) simplifies
to the canonical, unit-mass Klein-Gordon equation
\be
\l{WDWcanonical}
\left[ \frac{\partial^2}{\partial u \partial v} -1 \right] \Phi =0 .
\ee

Hence, the $\{ u,v \}$ variables may
be viewed as  null coordinates over a region of
$(1+1)$-dimensional Minkowski space-time spanned by the space-like coordinate
$X=u+v$ and time-like coordinate $T=u-v$.
These null coordinates are restricted to lie in the range
$ (u,v) \in (0,+\infty )$, so $X\ge |T|$. We may therefore view
the wavefunction $\Phi$ as a classical particle of unit
mass that is confined   within the Rindler wedge of  $(1+1)$-dimensional
Minkowski space-time.

The semi-classical  limit of the vacuum models
may also be analyzed by introducing the variables (\ref{u}).
In the WKB approximation,
one  substitutes solutions of the  form $\Phi_{\rm WKB} \approx e^{-iS /\hbar}$
into the Wheeler-DeWitt equation and considers
the limit $\hbar \rightarrow 0$. Applying this substitution in Eq.
(\ref{WDWcanonical}) implies that
the Hamilton-Jacobi equation takes the form
$S_uS_v =-1$, where a subscript denotes partial differentiation.
One
solution to this equation is $S=-i(cu+c^{-1}v)$, where $c$ is an arbitrary,
complex constant. However, the family of wavefunctions $\Phi_c = e^{-c u-v/c}$
are also exact solutions to the full Wheeler-DeWitt equation
(\ref{WDWcanonical}) \c{page}. Hence, the
WKB approximation is exact in this case.

$|\Phi_c |$ is bounded everywhere when ${\rm Re}$ $c\ge 0$
and the  wavefunction is exponentially damped
for large $\alpha$ if  ${\rm Im}$ $c=0$.
Unfortunately, however, it does  not satisfy the quantum wormhole boundary
conditions because
it  decays  too rapidly \c{HP1990}.
On the other hand, more general solutions to Eq. (\ref{WDWcanonical})  may
be generated in terms of linear superpositions of this
family. The class of wavefunctions $\Phi_c$ may therefore
be physically relevant. In general, the superpositions have the form
\be
\l{gensol}
\Phi = \int_C dc M(c) e^{-c u - v/c} ,
\ee
where $M(c)$ is an  arbitrary function of the parameter $c$ and
$C$ represents the contour of integration in the complex plane \c{page}.

Different solutions to Eq. (\ref{WDWcanonical})
correspond to different choices for $M(c)$ and $C$.
For example, if we specify
$M (c) =\frac{1}{2} c^{(i\epsilon -3)/3}$, where
$\epsilon $ is an arbitrary,  real constant,  and perform the integration
over the positive half of the real axis, we find that
\bea
\l{besselsolution}
\Phi_{\epsilon} =
\frac{1}{2} \int^{\infty}_0 dc c^{(i \epsilon-3)/3} e^{-c u- v/c} \nonumber
\\
= \left( \frac{v}{u}\right)^{i\epsilon /6} K_{i \epsilon/3} \left( 2\sqrt{uv}
\right) \nonumber \\
= K_{i\epsilon /3} \left[ \frac{1}{6} e^{2\alpha +2\beta_++2\sqrt{3}\beta_-}
\right] \exp \left[ -  \frac{i\epsilon}{6} \left( 8\alpha +2\beta_+
+2\sqrt{3}\beta_- \right) \right] ,
\eea
where $K$ is  the modified Bessel function of order $i\epsilon /3$ \c{page}.

The argument of this function is given by
\be
2\sqrt{uv} = |I| =\frac{1}{6} e^{2 \alpha +2 \beta_+ + 2\sqrt{3} \beta_-}
\ee
and corresponds to the high anisotropic limit of the Euclidean
action  (\ref{action}). It is also  directly proportional to the square
root of the superpotential (\ref{superpot}).
The modified Bessel function is exponentially damped for sufficiently
large arguments and the wavefunction takes the form
$\Phi_{\epsilon} \propto  e^{-|I|}$ for $\sqrt{uv} >|\epsilon | /6$. In this
region of minisuperspace, therefore, the wavefunction  reduces to
the Euclidean  form of  Eq. (\ref{action}).

The nature of solution (\ref{besselsolution}) is different for
smaller values of the scale factor. Indeed, it oscillates for
$0< |I| < |\epsilon | /3$, so the boundary
$I=|\epsilon | /3$ represents a point of maximum expansion. The
wavefunction has the asymptotic form $\Phi_{\epsilon}
\propto u^{-i\epsilon /3}$ as $
\alpha \rightarrow - \infty $ and therefore
represents plane waves in the minisuperspace variables when
$\epsilon \ne 0$. It
oscillates an infinite  number of times as the spatial metric degenerates and
this behaviour is interpreted as  a cosmological singularity.

Solution (\ref{gensol}) may provide insight into the quantum
nature of cosmological
singularities in  these highly anisotropic Universes.
The singularity arose in this case because a particular
superposition   of Euclidean solutions was considered.
In view of this, it is natural to investigate  whether the   singularity
may be eliminated by considering alternative superpositions.

In order to pursue this possibility further,
we introduce the new coordinate pair
\be
\l{airyvar}
\mu \equiv \frac{v}{2} +\sqrt{2u}, \qquad \nu \equiv \frac{v}{2} -
\sqrt{2u} .
\ee
This change of variables transforms  Eq. (\ref{WDWcanonical}) into
\be
\left[ \frac{\partial^2}{\partial \mu^2} -\frac{\partial^2}{\partial
\nu^2} - \mu +\nu \right] \Phi =0\
\ee
and the general, separable solution to this equation is given by
\begin{equation}
\label{airys}
\Phi_m = \left[ c_1 {\rm Ai} (m +\mu ) +
c_2 {\rm Bi} ( m+  \mu ) \right]
\left[ c_3 {\rm Ai} (m + \nu ) +c_4 {\rm Bi} (m +\nu )\right] ,
\end{equation}
where ${\rm Ai}(x)$ and ${\rm Bi}(x)$ are Airy functions and
$\{ m ,c_j \}$ are arbitrary constants.

This solution may be expressed in the form of
Eq. (\ref{gensol}) when  $c_j$ satisfy  appropriate conditions.
If we specify
\be
M(c) = \frac{\sqrt{2} i}{c^{3/2}} \exp \left[ \frac{2}{3c^3} -\frac{2m}{c}
\right] .
\ee
Eq. (\ref{gensol}) takes the form
\be
\l{HLintegral}
\Phi_m = \int_C \frac{d{\tilde{c}}}{{\tilde{c}}^{1/2}} \exp \left[ -
\frac{{\tilde{c}}^3}{12}
+(\mu +\nu +2m) \frac{{\tilde{c}}}{2} +\frac{1}{4{\tilde{c}}} (\mu -\nu )^2
\right] ,
\ee
where $\tilde{c} \equiv -2/c$.
Halliwell and Louko have shown how integrals of this form may be evaluated
\c{HL1989}.
Different choices for the contour $C$ result in different products
of Airy functions. Since we are interested in
superimposing bounded wavefunctions, we assume that Re $c>0$.
We therefore choose the contour of integration to lie to  the
left of the origin.  The result of the integration is
\be
\l{airy}
\Phi_m =  {\rm Ai} (\mu +m ) {\rm Ai} (\nu +m)
\ee
and, modulo a constant of proportionality,
this is equivalent to Eq. (\ref{airys})
with $c_2=c_4 =0$.

This solution has an interesting feature.
The Airy function ${\rm Ai}(x)$ is exponentially
damped for large positive arguments, but exhibits oscillatory
behaviour if $x<0$ \c{AS}.
However, the variable $(\mu +m )$ is  positive-definite for all values
of the scale factor if $m>0$.  In this case, it follows that
the wavefunction will  represent classically forbidden
geometries when $(\nu + m) >0$, but will correspond to Lorentzian solutions
when $(\nu +m) <0$. Now, when  the spatial volume of the
Universe becomes vanishingly small $(\alpha \rightarrow -\infty )$,
$u \rightarrow 0$ and $v \rightarrow
+\infty$, so $\mu \rightarrow \nu \rightarrow +\infty$.
Hence, the wavefunction vanishes, but  does not oscillate,
as the spatial metric degenerates.
Consequently, there is no singular behaviour at the origin. As the
scale factor grows, however, $(\nu +m)$ decreases and eventually becomes
negative.  At this point the wavefunction begins to exhibit oscillatory
behaviour. In effect, the Universe tunnels from a classically
forbidden regime into a Lorentzian domain when $\nu \approx -m$.

The above solutions were derived after the momentum constraints had been
reduced at the classical level.
However, it is more accurate to view these  constraints as
operators that annihilate the wavefunction of the
Universe. Consequently, it is better not to impose the
assumption of diagonality before quantization. Indeed, Ashtekar
and Samuel have argued that this restriction may
result in a loss of generality at the quantum level \c{AS1991}.
In the remainder  of this work, therefore,
 we shall consider the quantization procedure when the assumption
of diagonality is dropped.
In particular, we will investigate the vacuum Bianchi IX model.

The quantum constraints for this model were recently derived
by Higuchi and Wald \c{HW1995}.
We  briefly summarize  their main results here.
They parametrized each point in
minisuperspace in terms of an orthogonal matrix
${\rm {\bf O}}$ and the eigenvalues $\{ \lambda_1 ,\lambda_2 ,
\lambda_3 \}$ of the metric {\bf h}. These eigenvalues satisfy the conditions
$\lambda_1 \ge \lambda_2 \ge \lambda_3 >0$
and $\lambda_1 \lambda_2 \lambda_3 =1$. The  matrix ${\rm {\bf O}}$
rotates the eigenvectors  of the metric into a new orthonormal
basis.  It can be shown that the quantum momentum constraints
are satisfied  if the wavefunction is
invariant under the action of {\bf Oh}${\rm {\bf O}^{-1}}$. It must
therefore  be  symmetric  under permutations in $\lambda_i$.

If one chooses
$\lambda_1 \equiv e^{2(\beta_++\sqrt{3}\beta_-)}$,
$\lambda_2 \equiv e^{2(\beta_+-\sqrt{3}\beta_-)}$ and $\lambda_3\equiv
e^{-4\beta_+}$, the Wheeler-DeWitt equation may be written as
\be
\l{WDWmom}
\left[ \frac{\partial^2}{\partial \alpha^2} -\frac{1}{C(\beta_{\pm})}
\sum_{j=\pm} \frac{\partial}{\partial \beta_j} C(\beta_{\pm})
\frac{\partial}{\partial \beta_j} +U -90 \xi \right] \Psi =0 ,
\ee
where
\be
C(\beta_{\pm})  = 8\left| {\rm sinh} \left( 2\sqrt{3} \beta_- \right)
{\rm sinh}  \left(3\beta_+ -\sqrt{3}\beta_- \right)
{\rm sinh} \left( 3\beta_+ +\sqrt{3}
\beta_- \right) \right|
\ee
and $\xi$ is a numerical constant.
The non-trivial contribution from $C(\beta_{\pm})$ arises because
the volume element on superspace is affected by a term that depends on the
eigenvalues of the metric.

In this representation, the momentum constraints are satisfied
if the wavefunction is invariant under 120-degree rotations in the
$(\beta_+ ,\beta_- )$ plane and reflections in $\beta_-$.
An appropriate linear  combination of wavefunctions satisfying
these symmetry conditions
can always be constructed. Formally, one may write
the full wavefunction as
$\Psi = \Phi + R\Phi +R^{-1}\Phi +\Phi(\beta_-\rightarrow
-\beta_-)$, where $\Phi$ represents a particular  solution
to Eq. (\ref{WDWmom}) and $R$ is the 120-degree rotation matrix in the
$(\beta_+,\beta_-)$ plane \c{private}.
On a practical level, therefore,
one need only find solutions to the modified Wheeler-DeWitt
equation (\ref{WDWmom}).

However, it is not clear at present how
one might proceed to solve  this equation in full
generality. In view of this,
we shall consider the two limiting cases  where $\beta_{\pm}  \gg 1$ and
$\beta_+ \gg 1$, $|\beta_-| \ll 1$.

In the former case  $C(\beta_{\pm}) \approx e^{6\beta_+ +2\sqrt{3}\beta_- }$
and $U=48 uv$.
Eq. (\ref{WDWmom}) therefore
simplifies to\footnote{Note that we are assuming implicitly
that $\beta_+ \ge \beta_-$ in this analysis. This condition ensures that
$C(\beta_{\pm})$ has an approximately exponential form.}
\be
\l{quantum}
\left[ \frac{\partial^2}{\partial \alpha^2} -\frac{\partial^2}{\partial
\beta_+^2} -\frac{\partial^2}{\partial \beta_-^2} -6
\frac{\partial}{\partial \beta_+} -2\sqrt{3}
\frac{\partial}{\partial \beta_-} +U -90 \xi \right] \Psi =0.
\ee

A quantum wormhole solution to this equation may be found by
rescaling the wavefunction such that
$\varphi = \Psi e^{3\beta_+ + \sqrt{3}\beta_-}$.  This wavefunction
satisfies an equation that is formally equivalent
to Eq. (\ref{WDW1}), where  $q=0$ and $p^2 =90\xi -12$.
Hence, one solution is given by
$\varphi =e^{-2\sqrt{uv}}$  for the special case $\xi = 21/90$.

Further solutions may be found by defining a new wavefunction $\Phi$:
\be
\Phi (\alpha ,\beta_{\pm} ) = \Psi (\alpha ,\beta_{\pm} )
e^{-c_+\beta_+ +c_-\beta_-} ,
\ee
where
\bea
c_+\equiv \frac{\sqrt{3}}{2} \sqrt{ 12-90 \xi} -3 \nonumber \\
c_- \equiv \frac{1}{2} \sqrt{12-90\xi} +\sqrt{3} .
\eea
The Wheeler-DeWitt equation (\ref{quantum}) transforms to
\bea
\l{eq1}
\left[ \frac{\partial^2}{\partial \alpha^2} -\frac{\partial^2}{\partial
\beta_+^2} -\frac{\partial^2}{\partial \beta_-^2} -\sqrt{3} \left( 12
-90\xi \right)^{1/2}
\frac{\partial}{\partial \beta_+} +
\left( 12-90 \xi \right)^{1/2} \frac{\partial}{\partial \beta_-} \right.
\nonumber \\
\left.
+\frac{1}{3} e^{4\alpha +4\beta_+ +4\sqrt{3} \beta_-}  \right] \Phi =0
\eea
and this equation is formally equivalent to Eq. (\ref{eq}) with
$r=(12-90\xi)^{1/2}$. It therefore transforms
into the unit-mass Klein-Gordon equation (\ref{WDWcanonical})
when the null variables (\ref{u}) are introduced.
Hence, the above analysis and solutions
will also be relevant in this more general quantization
procedure
if the model is sufficiently anisotropic. The range of $\beta_{\pm}$ must be
such that $C(\beta_{\pm})$
can be viewed as a purely exponential function.

We will conclude by considering
the case where $\beta_+ \gg 1$ and $|\beta_- |\ll 1$.
It follows that $C(\beta_{\pm})
\approx  4\sqrt{3}|\beta_- |e^{6\beta_+}$
in this limit and the Wheeler-DeWitt equation (\ref{WDWmom})
therefore takes the approximate form
\be
\l{betaminus}
\left[ \frac{\partial^2}{\partial \alpha^2} -
\frac{\partial^2}{\partial \beta_+^2}-\frac{\partial^2}{\partial \beta_-^2}
 -\frac{1}{\beta_-} \frac{\partial}{\partial \beta_-} + (9-90\xi ) +
e^{4\alpha +4\beta_+} \beta_-^2 \right] \Theta =0 ,
\ee
where $\Theta \equiv e^{3\beta_+}\Psi$. Without loss of generality,
we have eliminated
the numerical constant in front of the superpotential via a linear shift in
$\alpha$.

To proceed, we change variables to the
null coordinates
\be
\rho \equiv \alpha + \beta_+, \qquad  \eta \equiv
\alpha - \beta_+ .
\ee
In this case, Eq. (\ref{betaminus}) transforms to
\be
\l{result}
\left[ 4\frac{\partial^2}{\partial \rho \partial \eta}
-\frac{\partial^2}{\partial \beta^2_-} -\frac{1}{\beta_-}
\frac{\partial}{\partial \beta_-} +(9-90\xi ) + \beta^2_- e^{4\rho} \right]
\Theta =0
\ee
and it follows that  the wavefunction $\Theta$
is an eigenstate of $\eta$. In order
to solve this equation,
  we assume that $\Theta$ has the generic form \c{KR1989}
\be
\Theta =\exp \left[ -iE\eta  -B(\rho) -\lambda A(\rho) \beta_-^2 \right] ,
\ee
where $A$ and $B$ are arbitrary   functions of the null variable $\rho$
and $\{ \lambda,  E \}$ are arbitrary constants. After substitution of
this ansatz, Eq.
(\ref{result}) separates into
two, first-order ordinary differential equations:
\be
\l{other}
4iEB' +4\lambda A +9-90\xi =0
\ee
\be
\l{riccati}
4iE\lambda A' -4\lambda^2 A^2 +e^{4\rho} =0 ,
\ee
where a prime denotes differentiation with respect to $\rho$.

If we introduce a new function $D(\rho)$, where $A(\rho) \equiv
d \ln D/d\rho$,  and choose $\lambda =-iE$,
Eq. (\ref{riccati}) simplifies to
\be
D'' +\frac{1}{4E^2}e^{4\rho} D =0 .
\ee
The general solution to this equation is given by
\be
\l{D}
D=Z_0 \left[ \frac{1}{4E}e^{2( \alpha +\beta_+ )}\right] ,
\ee
where $Z_0$ is an arbitrary,  linear combination of ordinary, zero-order
Bessel functions. Finally, Eq. (\ref{other}) may now be
solved by direct integration after substitution of Eq.  (\ref{D}). We conclude,
therefore,
that one wavefunction satisfying Eq. (\ref{betaminus}) is
\be
\Psi = \left[ Z_0 \left( e^{2\rho}/4E\right) \right]^{-1}
\exp \left[ iE (A\beta_-^2 -\eta )
- \frac{i}{E} \left( \frac{9-90\xi}{4} \right)
\rho  -3\beta_+ \right] .
\ee

To summarize, we have considered the quantization of the homogeneous
Bianchi class A cosmologies in the limit of high anisotropy.
A number of exact solutions to the quantum constraints were found.
We
considered first the  approach whereby the momentum constraints are reduced
at the classical level prior to quantization.
Wavefunctions that represent tunnelling from a classically forbidden
state were presented. These solutions may be expressed as a linear
superposition of purely  Euclidean wavefunctions.
We then considered a quantum reduction of the momentum constraints for
the vacuum Bianchi IX cosmology. The modified Wheeler-DeWitt equation
may be solved if appropriate conditions are satisfied.
These solutions  should prove useful
for investigating some of the fundamental questions that arise in quantum
cosmology.

\vspace{.2in}

We would like to thank A. Higuchi for helpful  communications on how
the momentum constraints can be satisfied at the quantum level. This
work was supported by the Particle Physics and Astronomy Research Council
(PPARC), UK.

\vspace{.2in}

\frenchspacing
\def\prl#1#2#3{{ Phys. Rev. Lett.} {\bf #1}, #2 (#3)}
\def\prd#1#2#3{{ Phys. Rev. D} {\bf #1}, #2 (#3)}
\def\plb#1#2#3{{ Phys. Lett. B} {\bf #1}, #2 (#3)}
\def\npb#1#2#3{{ Nucl. Phys. B} {\bf #1}, #2 (#3)}
\def\apj#1#2#3{{ Ap. J.} {\bf #1}, #2 (#3)}
\def\apjl#1#2#3{{ Ap. J. Lett.} {\bf #1}, #2 (#3)}
\def\cqg#1#2#3{{Class. Quantum Grav.} {\bf #1}, #2 (#3)}
\def\grg#1#2#3{{Gen. Rel. Grav.} {\bf #1}, #2 (#3)}
\def\mnras#1#2#3{{Mon. Not. R. astron. Soc.} {\bf #1}, #2 (#3)}


\end{document}